\documentclass[showpacs,preprintnumbers]{revtex4}
\usepackage{times}
\usepackage{bm}
\usepackage[]{graphicx}
\begin{document}



\title[Electron-molecules in quantum dots]
{Transitions between electron-molecule states in
electrostatic quantum dots}


\author{P. A. Maksym}
\affiliation{Department of Physics and Astronomy, University of Leicester, 
Leicester LE1 7RH, UK}
\author{H. Aoki}
\affiliation{Department of Physics, 
University of Tokyo, Hongo, Tokyo 113-0033, Japan}
\date{\today}
\begin{abstract}
Intermediate spin states that occur in electrostatic dots in the magnetic 
field regime just beyond the maximum density droplet are
investigated. The 5-electron system is studied with exact
diagonalization and group theory. The results indicate that the
intermediate spin states are mixed symmetry states with a superposition
of 5- and 4-fold electron-molecule configurations. 
A superposition of 5- and 4-fold correlation functions is found to 
reproduce the exact mixed symmetry pair correlation function to around
2\%.
\end{abstract}

\maketitle                   





\section{Introduction}

Recently, Nishi {\it et al} \cite{Nishi06} have identified the ground 
state quantum numbers
of a few-electron quantum dot in the strong magnetic field regime where the 
filling factor, $\nu$, is less than one. The device they studied is a
vertical pillar dot \cite{Kouwenhoven01} containing up to 5 electrons and
they identified the quantum numbers by comparing experimental excitation
spectra with those from an accurate theoretical model \cite{Maksym06}. 
In the particular
case of a 5-electron dot, the orbital angular momentum and spin, $(L, S)$, 
for the maximum density droplet state \cite{Kouwenhoven01} which occurs at 
$\nu = 1$ is $(10, 5/2)$. Beyond this, a sequence of transitions
occurs as $\nu$ decreases and the sequence of observed $(L,S)$ values is 
$(14,3/2)$, $(15,5/2)$, $(18,3/2)$ and $(20,5/2)$. However the work of 
Nishi {\it et al} only gives the ground state quantum numbers and further
theoretical analysis is needed to gain insight into the form of the ground
state. This is done in the present work.

The fact that the observed combinations of $L$ and $S$ are severely
restricted suggests that the ground states have a molecular form 
\cite{Maksym00}. That is, the electron ground state is localised around the
classical equilibrium positions and corresponds to rotational and
vibrational motion. This turns out to be the case for the fully
spin-polarised states $(S=5/2)$ but the intermediate spin states $(S=3/2)$
are different. In this case, states centred on different classical minima
are allowed to couple and the ground state is an unusual state of mixed
symmetry. The evidence for this comes from the form of the ground state 
pair correlation functions which are calculated by exact diagonalization 
(section \ref{diagsect}) and from group theory which provides the link 
between the ground state quantum numbers and the point symmetry of the 
electron molecule (section \ref{groupsect}). In addition, it turns out that
the pair correlation function of the mixed symmetry states is
well-approximated by a superposition of pair correlation functions for each
symmetry type (section \ref{groupsect}).

\section{Pair correlation functions}
\label{diagsect}

The exact diagonalization calculations are performed with the same
theoretical model that Nishi {\it et al} used to analyse their data 
\cite{Maksym06}. The model includes the effect of electron-electron
interactions, screening, finite dot thickness and image charges.
Physically, this leads to a model in which
electrons are confined in a parabolic potential, move in two
dimensions and interact via a modified interaction. The confinement energy, 
$\hbar\omega$, for the 
5-electron dot is 4.88 meV. The electron ground states are
found by diagonalization of the Hamiltonian in a Fock-Darwin basis that
includes higher Landau levels and the numerical relative error in the 
ground state energies is about $3 \times 10^{-4}$. The calculations
reproduce experimental addition energies to about 5\% \cite{Nishi06}. 

\begin{figure}
\includegraphics[width=3.5cm]{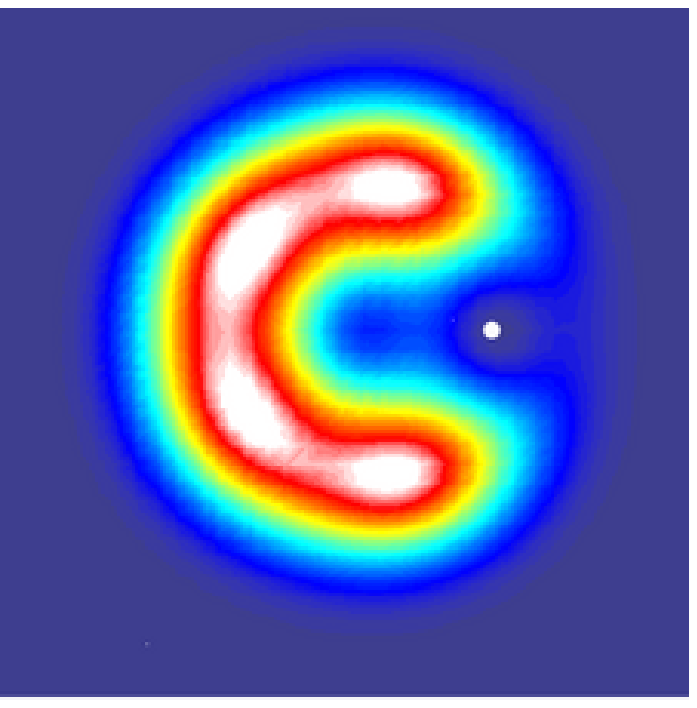}
\includegraphics[width=3.5cm]{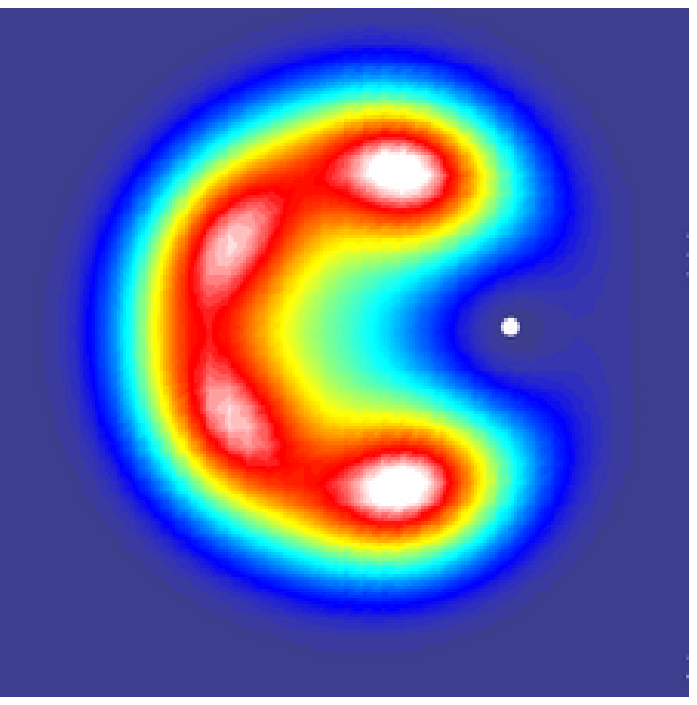}
\includegraphics[width=3.5cm]{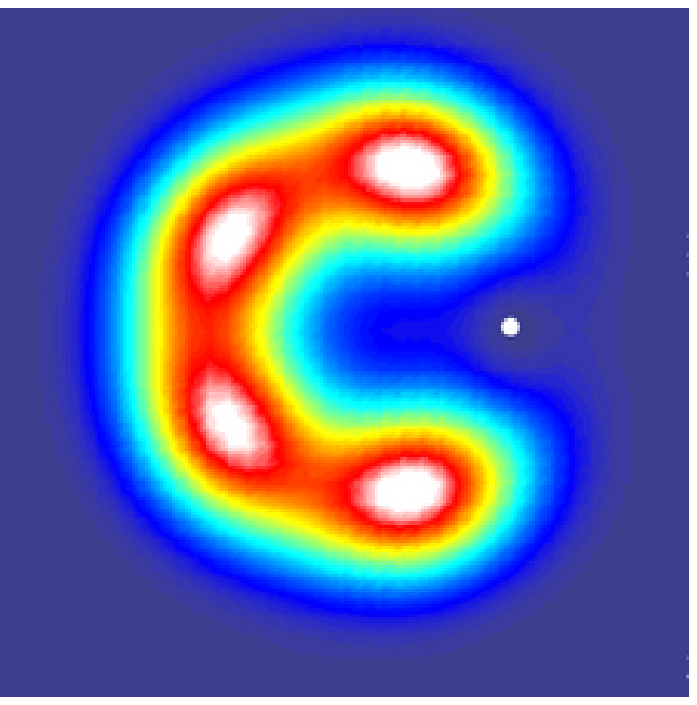}
\caption{Spin summed pair correlation functions for $(L,S)=(15,5/2)$ (left), 
$(18,3/2)$ (centre) and $(20,5/2)$ (right). The white spot indicates 
${\bm r}_0$.
} \label{ptotfig}
\end{figure}
\begin{figure}
\includegraphics[width=3.5cm]{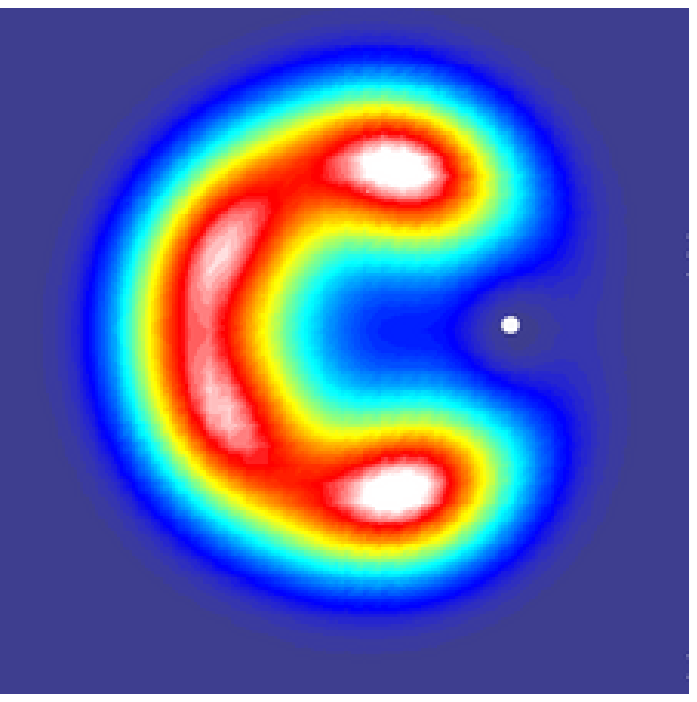}
\includegraphics[width=3.5cm]{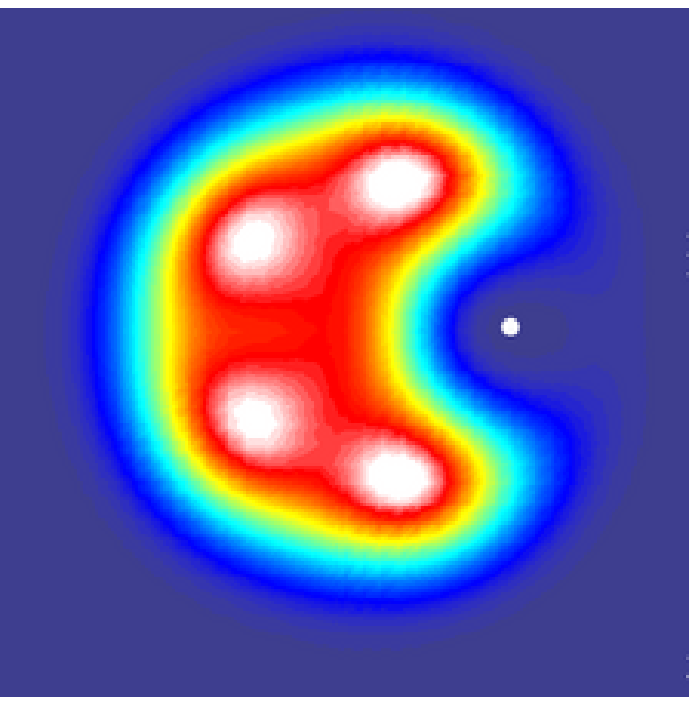}
\caption{Spin resolved pair correlation functions for $(L,S)=(18,3/2)$:
like spin (left), unlike spin (right).
} \label{pspinfig}
\end{figure}
The pair correlation function is the probability of
finding an electron of spin $s$ at ${\bm r}$ given that there is one of
spin $s'$ at ${\bm r}_0$:
\begin{equation}
P_{ss'}({\bm r},{\bm r}_0) = \frac {(2 \pi \lambda^2)^2}{N(N-1)}\langle
\sum_{i\ne j} \delta({\bm r}_i - {\bm r}) \delta_{ss_i}
\delta({\bm r}_j - {\bm r}_0)  \delta_{s's_j}
\rangle,
\label{pdefinition}
\end{equation}
where the angle brackets denote a quantum mechanical expectation value
and $N$ is the number of electrons \cite{Maksym96}. 
$\lambda$ is the
length parameter for harmonic oscillator confinement in a magnetic field,
$\lambda^2 = \hbar / 2 m^* \Omega$ where $\Omega^2 = \omega^2 + \omega_c^2$
and $\omega_c$ is the cyclotron frequency. 

Spin summed pair correlation functions for the 
$(15,5/2)$, $(18,3/2)$ and $(20,5/2)$ states are shown in 
Fig.~\ref{ptotfig}. The white regions
correspond to the largest values and each square corresponds to an area
of $12\lambda \times 12\lambda$ centred on the origin. In each case the 
electron density has the form of a ring with a minimum in the centre of the
dot. The fixed position ${\bm r}_0$ is taken to be on $x$-axis, at the
position of maximum density. The pair correlation functions for spin
polarised states show pentagonal symmetry: 4 peaks on a ring are
clearly visible in the figure and these peaks together with ${\bm r}_0$
form a pentagon. In contrast, the structure around the ring in the
intermediate spin case is significantly weaker than for the $(15,5/2)$ and 
$(20,5/2)$ cases. In addition, there is pronounced intensity in the
centre and the value at the origin is about 3 times larger than for the 
$(20,5/2)$ case.

Figure~\ref{pspinfig} shows spin resolved pair correlation functions for
the $(18,3/2)$ state. The like spin correlation function has a ring-like
structure but the symmetry is not 5-fold because there is a broad 
ridge
on the left side of the ring instead of two pronounced peaks. The unlike
spin correlation function does have 4 distinct peaks on the ring but the
peaks are superimposed on a plateau that extends into the centre of the
correlation function, which suggests an antiferromagnetic correlation 
between outer and central sites. 

In the strong magnetic field limit, the structure in the pair correlation 
functions tends to be associated with minima in the potential energy
\cite{Maksym96}. The global energy minimum configuration of 5 confined,
interacting 
electrons has the form of a pentagon but there is also a local minimum
about 0.5-1\% higher in energy which has the form of a square with one 
electron at its centre.
The form of the pair correlation functions for $(18,3/2)$ is consistent 
with the idea
that both configurations contribute to the quantum state. The
broad ridge on the left side of the ring in the like
spin correlation function is consistent with the overlap of peaks from 
5- and 4-fold configurations and the relatively large value at the centre 
of the unlike spin correlation function is consistent with a
contribution from the central electron in the 4-fold configuration. 

\section{Mixed symmetry states}
\label{groupsect}

The allowed $(L,S)$ values for rotational-vibrational states localised 
around potential energy
minima can be determined with group theory and this supports the idea that
the $(18,3/2)$ state is a mixture of symmetry types. The electron positions
that minimise the energy have either 5- or 4-fold symmetry. 
When the confinement is harmonic,
the centre of mass motion separates and the relative Hamiltonian can be
expresssed in the Eckart frame in which the rotational and vibrational
motion is approximately decoupled \cite{Maksym00}. The relative
Hamiltonian is conveniently expressed in terms of $2 N - 3$ 
normal co-ordinates $Q_i$ and one Euler angle $\chi$. Before 
antisymmetrization, the eigenstates have the form
\begin{eqnarray}
\Psi = \psi_{CM} \exp(-iL_{RM} \chi)
f_{L_{RM},n_1...n_{2N-3}}(Q_1,...,Q_{2N-3}) \psi_{spin}(S_z),
\label{vibestate}
\end{eqnarray}
where $\psi_{CM}$ is the centre of mass eigenstate, the exponential factor
is an eigenstate of relative angular momentum, $f$ is a vibrational
eigenstate, $n_i$ is the number of quanta in the $i$\/th vibrational mode
and $\psi_{spin}$ is the spin state. $L_{RM}$ is the relative
angular momentum quantum number. For ground states the
centre of mass angular momentum is 0 so $L_{RM} = L$. The electron states
are found by anti-symmetrizing $\Psi$. Application of the
antisymmetrization operator $\hat{A}$ to $\Psi$ either gives a valid 
electron state or zero, depending on whether $L$, $S_z$ and $n_i$ are
compatible with the Pauli principle. For ground states
$n_i=0$ so the antisymmetrization process gives the allowed values of 
$(L,S)$.

Although $\hat{A}$ generates a sum over $N!$ permutations, it is not
necessary to consider the full permutation group to find the allowed
$(L,S)$ values. It turns out that a cyclic permutation operating on $\Psi$
is equivalent to a rotation. This enables the $(L,S)$ values for a system
with $m$-fold symmetry to be found by considering $C_m$, the subgroup of the
permutation group that is isomorphic to the cyclic group of order $m$. The
relevant procedure is outlined in references \cite{Maksym00} and 
\cite{Maksym96} and examples are given for the spin polarised case. A
few additional points, relevant to the intermediate spin case,
are summarised here. The $N$-electron spin functions used in the procedure 
are eigenfunctions of the cylic permutations. For each value of $S_z$, the
entire set of $N$-electron spin functions forms a basis for a reducible
representation of $C_m$. Reduction of this representation gives the number
of spin functions associated with each representation of $C_m$ at a given
value of $S_z$. But what is needed is the number of spin functions for each
value of {\it total spin} $S$. This is found by an algorithm that depends
on use
of spin lowering operators. When $S_z$ takes its maximum value, $S^*_z=5/2$ 
in the present case, the only possible value of $S$ is $S=S_z$. Application 
of the spin lowering operator to $|S, S_z\rangle = |S^*, S^*\rangle$ generates 
$|S^*,S^*-1\rangle$. This enables all the states with $S=S^*$, $S_z=S^*$ and
$S=S^*$, $S_z=S^*-1$ to be
eliminated. In the remaining states, all those that have $S_z=S^*-1$ also
have $S=S^*-1$. Recursive application of the spin lowering operator
together with consideration of rotational symmetry therefore enables the
allowed $(L,S)$ values to be determined.

The group theoretical analysis shows that for states with 
$(L,S)=(5k,5/2)$, $(k{\rm: an~integer})$, 5-fold ground states are allowed 
and 4-fold states are forbidden. Hence a ground state localised around the 
5-fold global potential minimum is allowed.
In contrast, both 5-fold and 4-fold states are allowed at 
$(4k+2,3/2)$. The 5-fold energy minimum still favours a 5-fold state but if
the global minimum is close in energy to the local 4-fold minimum and the
coupling between the 5- and 4-fold states is sufficiently strong the ground
state is expected to be a mixture of the two symmetry types. The form of
the pair correlation functions suggests this is indeed happening.  

\begin{figure}
\includegraphics[width=3.5cm]{pcfj18totcs.eps}
\includegraphics[width=3.5cm]{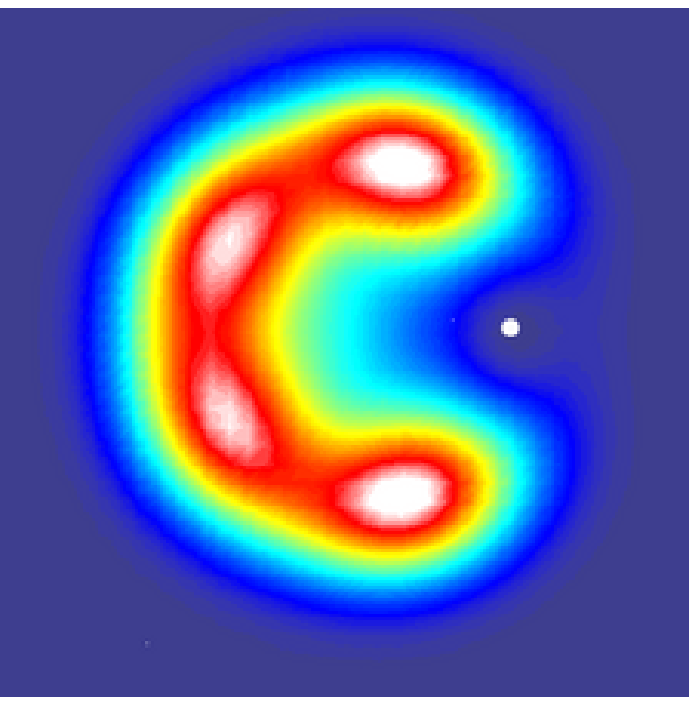}
\includegraphics[width=3.5cm]{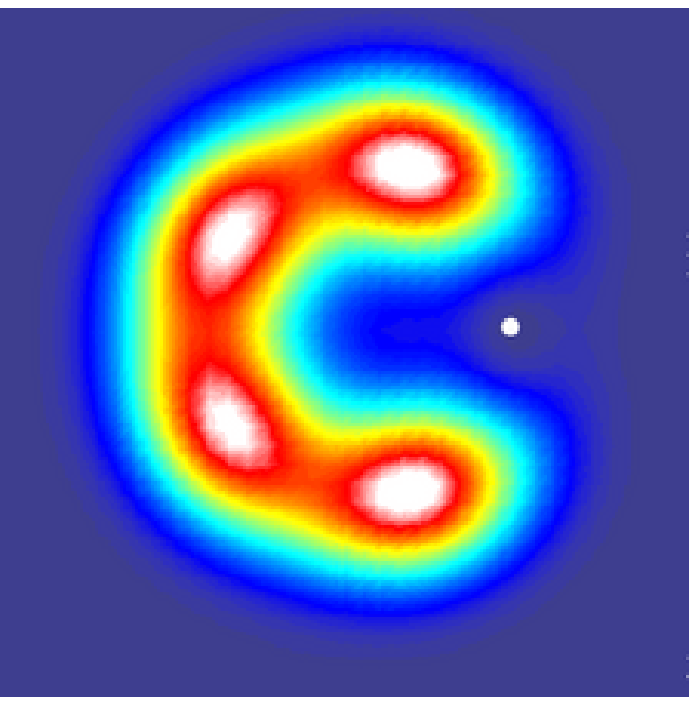}
\includegraphics[width=3.5cm]{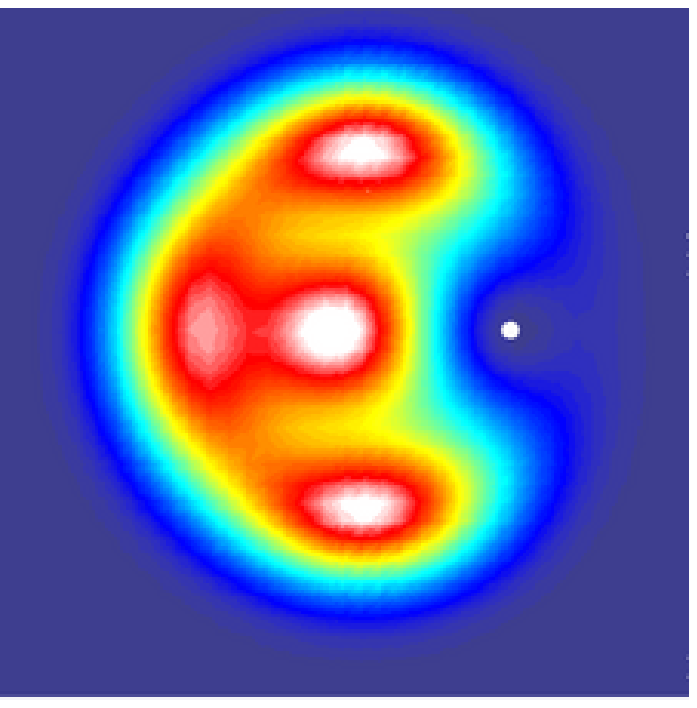}
\caption{Exact (left) and fitted (second left) correlation functions for 
$(L,S)=(18,3/2)$ together with 4- (right) and 5-fold (second right) components.
} \label{fitfig}
\end{figure}
To test this idea quantitatively, the pair correlation function for a
superposition of 5- and 4-fold states is computed. The $(18,3/2)$ state is
taken to have the form $\alpha |5\rangle + \beta |4\rangle$ where 
$|m\rangle$ is a $m$-fold state. The pair correlation function is 
$|\alpha|^2\langle5|\hat{P}|5\rangle + |\beta|^2\langle4|\hat{P}|4\rangle + 
2\Re(\alpha^*\beta
\langle5|\hat{P}|4\rangle)$, where $\hat{P}$ is the operator in 
Eq.~(\ref{pdefinition}).
If the cross term is assumed to be small this simplifies to 
$p\langle 5|\hat{P}|5\rangle + (1-p)\langle 4|\hat{P}|4\rangle$ where 
$p$ is the probability of
finding the system in the 5-fold state. The 5- and 4-fold states are taken
to be $(20,5/2)$ and $(18,5/2)$, the states of pure symmetry that are
closest in energy to $(18,3/2)$. Although the pure symmetry states have 
$S=5/2$, the {\it spatial} factors in these states will be similar to those
in $(18,3/2)$ if Eq.~(\ref{vibestate}) is a good approximation. $p$ is found
by fitting to the exact $(18,3/2)$ correlation function. $p=0.72\pm0.01$
fits the exact function with an RMS error of $4.9\times 10^{-4}$, about 2\%
of the typical peak height. The fact that the
5-fold component occurs with the largest probability is consistent with the
5-fold global minimum. Figure~\ref{fitfig} shows the exact correlation
function, together with the fitted one and the two components used for the
fit. In each case ${\bm r}_0$ is on the density maximum for 
$(18,3/2)$. This differs from the density maximum position for
$(18,5/2)$ by about 10\%. Consequently, the 4-fold component pair function 
has a small deviation from 4-fold symmetry. This
contributes to the 2\% error but it is clear that the main features of the
exact correlation function are well reproduced in the fit. 

The present analysis depends on the localised states,
Eq.~(\ref{vibestate}),
being a good approximation. In other words, the electron molecule theory
should describe physics. The theory is known to reproduce ground state
energies to about 1 part in $10^4$ in the very strong field regime where
the ground state angular momentum is large 
\cite{Maksym00,Maksym96,Imamura98}. When $\nu\sim 1$, it is still accurate 
to a few \%, sufficient for the present analysis.

\section{Discussion}

Both group theoretical analysis and the form of exact pair correlation
functions suggest that the intermediate spin states that occur just beyond
the maximum density droplet are mixed symmetry states in which two
different point symmetries coexist. For systems with {\it large numbers of 
electrons}, the regime beyond the maximum density droplet has previously been 
described in terms of spin textured states \cite{Oaknin96}. Large systems
can be divided into a core region and an edge region but in contrast, 
the present few-electron system is too small to have a core. Hence its states
resemble those of a molecule rather than those of a bulk system and have the
unique feature of mixed symmetry. The 5-electron system is the 
smallest for
which competing potential minima occur, hence the smallest in which mixed
symmetry states can occur. However competing minima are a common feature of
potential landscapes so it is likely that similar states will occur just
beyond the maximum density droplet in larger but still molecular systems.  


We thank Y. Nishi for suggesting the fitting procedure.

\end{document}